\documentclass[aps,pra]{revtex4}%
\usepackage{amsfonts}
\usepackage{amsmath}
\usepackage{amssymb}
\usepackage{graphicx}%
\providecommand{\U}[1]{\protect\rule{.1in}{.1in}}
\begin{document}
\preprint{ }
\title{Fermion-mediated interactions in a dilute Bose-Einstein condensate}
\author{D. H. Santamore$^{1,}$$^{2}$}
\author{Eddy Timmermans$^{3}$}
\affiliation{$^{1}$T-4, Theoretical Division, Los Alamos National Laboratory, Los Alamos,
New Mexico 87545}
\affiliation{$^{2}$Department of Physics, Temple University, Philadelphia, Pennsylvania 19122}
\affiliation{$^{3}$T-CNLS, Theoretical Division, Los Alamos National Laboratory, Los
Alamos, New Mexico 87545}


\begin{abstract}
We develop a diagrammatic perturbation treatment to calculate the
zero-temperature equation of state of the dilute gas mixture of a single spin
component Bose-Einstein condensate (BEC) and a normal Fermi gas of
indistinguishable (single spin) fermion particles. We find that the mean-field
description breaks down near the mechanical instability related to the phase
separation phenomenon. Our analysis shows that the instability is caused by
the competition of the usual short-range and fermion-mediated boson-boson
interactions, which result in a boson compressibility that diverges. In the
low BEC-density limit, we show that the diagrammatic analysis simplifies, we
sum part of the higher order diagrams, and we discuss the effects of other
higher-order contributions.

\end{abstract}
\date{\today }
\maketitle

\section{Introduction}

Cold atom technology provides a novel laboratory for the many-body study of
quantum liquids. In this paper, we focus on the low-temperature cold atom
mixtures of dilute boson and fermion quantum liquids. The simultaneous
trapping and cooling of fermionic and bosonic atoms has become a routine
ingredient of the experimental cold atom repertoire. The original motivation
for trapping these quantum gas mixtures was sympathetic cooling
\cite{Kett02,Hulet03,Salomon01}: In the quantum degenerate regime, the
standard evaporative cooling technique--the removal of the most energetic
particles--is more efficient with bosonic than with fermionic atoms.
Evaporatively cooled bosons, often a Bose-Einstein condensate (BEC), can
subsequently cool the fermion particles by thermal contact.

In addition to its sympathetic cooling use, theorists also have pointed out
intriguing prospects of cold atom technology for many-body studies: As the
elementary BEC excitations are acoustic phonon modes, fermion-BEC mixtures
consist of fermions interacting with a phonon fluid, thereby representing a
large class of systems. A cold atom feature that is highly unusual in this
broad context is the continuous tunability of the fermion-boson (and hence
fermion-phonon) interactions, obtained by adjusting the strength of a
homogeneous, external magnetic field--the Feshbach resonance. In general, the
unprecedented accessibility, novel knobs, and probes suggest that in their
role of prototypes of fermion-phonon fluid, the cold atom fermion-BEC mixtures
can be used to break new ground by exploring polaron-like self-localization
\cite{Fernando} or by measuring the rate of heat-exchange between strongly
coupled fermion-phonon fluids (temperature relaxation).

From the perspective of quantum liquid physics, a particularly relevant
phenomenon is the phase-separation transition that the fermion-BEC mixture is
expected to undergo as the fermion-boson interaction strength is increased
\cite{Viverit}. This transition is the analogue of the phase separation of
condensed $^{3}\mathrm{He}-^{4}\mathrm{He}$ fluid mixtures studied in
traditional low-temperature physics \cite{sep}. As we describe in Sec.
\ref{Sec_MFT}, we expect the cold atom phase-separation transition to be first
order \cite{Dima}, as was observed in the condensed-matter helium mixtures.
Accessing this transition by varying the densities and/or the interaction
strengths would allow the unusual, though highly relevant, experimental
exploration of first-order (zero temperature) quantum phase transitions. As a
first-order transition, its dynamics can involve nucleation or a spinodal
decomposition. At sufficiently low temperatures, the nucleation process cannot
be triggered by thermal activation and has to proceed via a many-body
tunneling process. Hence, the cold atom fermion-BEC mixtures promise the
prospect of many-body tunneling studies, as well as investigations of
first-order quantum phase transitions \cite{Lifshitz}.

The focus of our work in this paper is the role of mediated interactions--the
modification of the interactions of particles in one fluid caused by the
presence of the other fluid. This phenomenon is of fundamental importance:
fermionic particles interact via boson-field mediated interactions in the
standard model. In the phase separation of low-temperature condensed helium
mixtures, mediated interactions play a crucial role \cite{Bardeen}. In those
systems, the condensed $^{3}\mathrm{He}$ fermion liquid remains
\textquotedblleft normal\textquotedblright\ (i.e., does not become superfluid)
in the experimentally accessible regime, whereas the $^{4}\mathrm{He}$ boson
fluid can take on a BEC-like superfluid state. Boson mediated interactions are
expected to make the fermions superfluid \cite{Bardeen2} as well, but at
temperatures that are an order of magnitude lower than can be reached with
present cooling technology. The condensed $^{3}\mathrm{He}-^{4}\mathrm{He}$
mixtures undergo phase separation when the fraction of fermion particles
exceeds a minimal value: the system then breaks translational symmetry by
forming local regions in which only fermion particles reside. While our
understanding of theses systems remains limited by the difficulties involved
in accounting for strong interaction effects, theorists understood early on
that the phase-separation phenomenon is triggered by mediated interactions
\cite{Bardeen}. Mean-field studies of the cold atom fermion-BEC mixtures
indicated that these gas mixtures would similarly undergo phase separation
under experimentally attainable conditions. Hence, it should be possible to
study mediated interactions in a cold atom environment that is amenable to a
first-principle description.

Mediated interactions stem from a particular type of correlations: the local
density of the second fluid is altered near the position of a particle
\textquotedblleft1\textquotedblright\ of the first fluid, thereby changing the
potential energy experienced by another, nearby particle \textquotedblleft%
2\textquotedblright\ of the first fluid. However, most cold atom phase
separation studies have been carried out in the mean-field approximation,
which neglects all correlations. Other studies neglected dominant
contributions \cite{Wilkens02,Wilkens03,Viverit2}. The zero-temperature
equation of state and, hence, the density profile calculated within the
mean-field equation of state cannot be expected to give a quantitatively
correct description.

In Sec.\ \ref{Sec_MFT}, we derive the equation of the state of the above
system within the mean-field approximation. In Sec.\ \ref{Sec_linearresponse},
we describe the fermion-mediated boson-boson interactions starting from
linear-response theory. Then, we develop a general perturbation formalism that
accounts for the relevant correlations in the following four sections,
Secs.\ \ref{Sec_FmedBB}--\ref{Sec_sum}. Our second-order calculation indicates
that the instability associated with the spinodal decomposition is not a
saddle point instability, as predicted by the mean-field model, but is caused
by the divergence of the BEC compressibility. In the low-density BEC limit, we
find that the perturbation treatment simplifies and we show that higher-order
terms significantly lower the fermion density of the instability point and
make the instability line dependent upon the BEC density. From a partial
summation of diagrams, we find an analytical expression for the equation of
state and estimate the effect of other contributions.

\section{Cold Atom Background and Mean-Field Description\label{Sec_MFT}}

In this section, we describe the ground state of a dilute gas mixture of
single-component fermionic atoms and single spin projection bosonic atoms.
These fermions (bosons), interacting via short-range atom-atom interaction
potentials, are indistinguishable particles with mass $m_{F}$ ($m_{B}$),
distributed in space with average particle density $\rho_{F}^{0}\left(
\mathbf{r}\right)  $ $\left[  \rho_{B}^{0}\left(  \mathbf{r}\right)  \right]
$. We assume that the fermions are not paired into a superfluid, which may
require the system to have a temperature above the critical temperature
$T_{c}$ for mediated interaction induced fermion pairing
\cite{Pinesfirst,Cohen}. In practice, however, the $T_{c}$ for such fermion
pairing \cite{Eftemov,Bulgac} is much lower than any other relevant energy
scale so that there is a temperature regime in which for all practical
purposes we can treat the unpaired fermion-BEC mixture as a zero-temperature
system with a normal fermion fluid. Whether the $p$-wave pairing can be
achieved may also depend on other factors: for instance, the inevitable loss
processes that give a finite lifetime to the cold atom trap systems also heat
up the system \cite{Timmermans}. This is a process that will compete with the
formation of any ultralow-temperature phase of matter.

As mentioned before, the mean-field description commonly used in cold atom
physics neglects correlations and fluctuations in the expression of all
expectation values. Accordingly, the interaction energy of this system,
$E_{\mathrm{int}}$, is approximated as
\begin{equation}
E_{\mathrm{int}}=\lambda_{BF}\int d^{3}r\;\rho_{F}^{0}\left(  \mathbf{r}%
\right)  \rho_{B}^{0}\left(  \mathbf{r}\right)  \mathbf{+}\frac{\lambda_{BB}%
}{2}\int d^{3}r\;\rho_{B}^{0}\left(  \mathbf{r}\right)  \rho_{B}^{0}\left(
\mathbf{r}\right)  ,\label{MF1}%
\end{equation}
where $\lambda_{BF}$ and $\lambda_{BB}$ are the interaction strengths with
$\lambda_{BF}=2\pi\hbar^{2}(m_{F}^{-1}+m_{B}^{-1})a_{BF}$ and $\lambda
_{BB}=\left(  4\pi\hbar^{2}/m_{B}\right)  a_{BB}$, where $a_{BF}$ and $a_{BB}$
are the fermion-boson and boson-boson scattering lengths, respectively. We
assume that the effective boson-boson interactions are repulsive, i.e.,
$a_{BB}>0$, in order to have the system be mechanically stable.

M\"{o}lmer \cite{Molmer} pointed out that the trapped boson-fermion quantum
gas mixtures can take on distinct spatial arrangements at low temperatures.
When the fermion and boson particles mutually repel each other, $\lambda
_{BF}>0$, the mixtures can take on spatial configurations in which one region
of space is occupied by only fermion particles, an arrangement akin to that of
the phase-separated state in condensed $^{3}\mathrm{He}-^{4}\mathrm{He}$
mixtures. On the other hand, when the fermions and bosons attract each other,
$\lambda_{BF}<0$, the mutual attraction causes a density increase in the
regions of fermion-boson overlap. If the attraction is sufficiently strong,
the density increase leads to a collapsing instability \cite{Roth} that was
observed experimentally by Inguscio and co-workers \cite{collapse}.

In the mixtures of repulsive fermion-boson interaction, Viverit \textit{et
al.} \cite{Viverit} calculated the zero-temperature phase diagram in the
mean-field approximation. Specifically, they discovered that the fermion-boson
densities for an initially homogeneous mixture would take one of the following
phases: (a) remain homogeneously mixed (phase I), (b) spatially separate into
spatial regions of all fermion-gas and spatial regions containing a
fermion-BEC mixture (phase II), and (c) separate into fermion-only and
BEC-only regions (phase III). They have determined the density regime
boundaries of the phase diagram by equating the mean-field pressures as well
as the mean-field chemical potentials of those particles that reside in both
kinds of spatial regions. The pressures and chemical potentials follow from
the zero-temperature \textquotedblleft equation of state\textquotedblright:
the ground-state energy of a homogeneous mixture.

The mean-field equation of state of a homogeneous mixture of $N_{F}$ fermions
and $N_{B}$ bosons confined to a macroscopic volume $\Omega$, corresponding to
average fermion and boson densities $\rho_{F}^{0}=N_{F}/\Omega$ and $\rho
_{B}^{0}=N_{B}/\Omega$, respectively, is then equal to
\begin{equation}
E=\frac{\lambda_{B}N_{B}^{2}}{2\Omega}+\frac{\lambda_{BF}N_{F}N_{B}}{\Omega
}+\frac{3}{5}N_{F}\frac{\hbar^{2}k_{F}^{2}}{2m_{F}},\label{MFEOS}%
\end{equation}
where $k_{F}$ is the Fermi momentum $k_{F}=\left(  6\pi^{2}\rho_{F}%
^{0}\right)  ^{1/3}$. Then, the fermion and boson chemical potentials,
$\mu_{B}=dE/dN_{B}$ and $\mu_{F}=dE/dN_{F}$, follow from Eq.~(\ref{MFEOS}),
\begin{align}
\mu_{F} &  =\frac{\hbar^{2}k_{F}^{2}}{2m_{F}}+\lambda_{BF}\rho_{B}%
^{0},\label{MFchempot}\\
\mu_{B} &  =\lambda_{BB}\rho_{B}^{0}+\lambda_{BF}\rho_{F}^{0}%
.\label{MBchempot}%
\end{align}
We obtain the same chemical potential equations by minimizing the free-energy
function $F=E-\mu_{B}N_{B}-\mu_{F}N_{F}$ with respect to $N_{F}$ and $N_{B}$.
Note that the homogeneous mixture represents the physical ground state only if
Eqs. \ (\ref{MFchempot}) and (\ref{MBchempot}) yield the global minimum of
$F$. This global minimum condition requires that the second derivatives of $F$
(or $E$) with respect to $N_{B}$ and $N_{F}$ be positive, i.e., $\partial
\mu_{B}/\partial\rho_{B}^{0}>0$ and $\partial\mu_{F}/\partial\rho_{F}^{0}>0$.
Physically, these mathematical conditions imply that the compressibility of
the fermion and boson systems should be positive, as the isothermal
compressibility, $\kappa_{j}$, is inversely proportional to the chemical
potential derivative,
\begin{equation}
\kappa_{j}=-\frac{1}{\Omega}\left(  \frac{\partial\Omega}{\partial P_{j}%
}\right)  _{T}=\left[  \left(  \rho_{j}^{0}\right)  ^{2}\frac{\partial\mu_{j}%
}{\partial\rho_{j}}\right]  ^{-1},
\end{equation}
where $P_{j}$ denotes the partial pressure experienced by gas $j$ ($j=B$ or
$j=F$), and where the subscript $T$ indicates that the pressure derivative
should be taken at constant temperature (zero temperature in this case). In
addition, the condition of a global minimum also implies that the extremum
cannot be a saddlepoint, which also requires $\left(  \delta\mu_{B}%
/\partial\rho_{F}^{0}\right)  ^{2}<\left(  \partial\mu_{B}/\partial\rho
_{B}^{0}\right)  \left(  \partial\mu_{F}/\partial\rho_{F}^{0}\right)  $.
Viverit \textit{et al}. \cite{Viverit} noted that when the fermion density
exceeds a critical value, $\rho_{F,\mathrm{crit}}$, where
\begin{equation}
\rho_{F,\mathrm{crit}}=\frac{3}{4\pi a_{BF}^{3}}\left[  \frac{a_{BB}/a_{BF}%
}{(1+m_{F}/m_{B})(1+m_{B}/m_{F})}\right]  ^{3},
\end{equation}
the extremum is a saddlepoint and the system can lower its energy by rolling
down the saddlepoint in different $N_{B}/N_{F}$ directions in different
spatial regions. The \textquotedblleft rolling down the
saddlepoint\textquotedblright\ suggests a mechanism by which the mixture can
phase-separate into regions of roughly equal size, i.e., undergo spinodal decomposition.

\begin{figure}[ptb]
\begin{center}
\includegraphics[height=2.4512in, width=3.3996in] {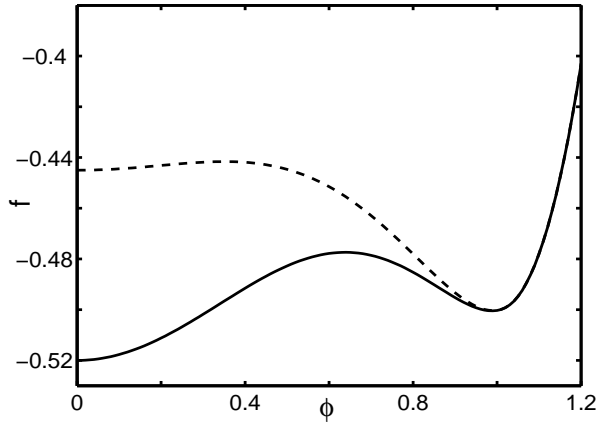}
\end{center}
\caption{Scaled free energy $f$ as a function of $\phi$ $\left[
c.f.Eq.\ (\ref{freeen})\right]  $. Dotted line shows the stable system (one
global minimum) and solid line shows the system that can undergo first-order
phase transition (one local minimum and one global minimum at the origin).}%
\label{freeE}%
\end{figure}

On the phase diagram in the boson fermion density plane, the line of spinodal
decomposition, $\rho_{F}=\rho_{F,crit}$, is located in the region with fermion
densities above the phase I/II boundary \cite{Viverit}. Thus, one might wonder
how the mixture phase-separates if the densities of the initial homogeneous
mixture were located in between the I/II boundary line and the line of
spinodal decomposition according to the mean-field description. We gain
insight into this dynamics by plotting the free energy density $\mathcal{F}%
\left(  \phi,\mu_{F},\mu_{B}\right)  =F/\Omega$ as a function of the
superfluid BEC order parameter, which, if all bosons are Bose-condensed, is
related to the boson density as $|\Phi|^{2}=\rho_{B}^{0}$. Actually, we find
it useful to scale both the BEC order parameter as well as the free energy
density. We introduce $|\Phi^{0}|$\ with $|\Phi^{0}|^{2}=\mu_{B}/\lambda_{BB}%
$, and set $\phi=\Phi/|\Phi^{0}|$. We also define the scaled free energy by
$f$, where $f=\mathcal{F}\lambda_{BB}/\mu_{B}^{2}$. Then, substituting the
fermion density in terms of the fermion and boson chemical potentials, we
obtain
\begin{equation}
f=\left\{
\begin{array}
[c]{c}%
-|\phi|^{2}+\frac{1}{2}|\phi|^{4}-\alpha_{\mathrm{MF}}\left[  1-\frac
{\left\vert \phi|^{2}\right\vert }{\beta_{MF}^{2}}\right]  ^{5/2},\text{
\ \ \ if }|\phi|<\beta_{\mathrm{MF}},\\
-|\phi|^{2}+\frac{1}{2}|\phi|^{4},\text{ \ \ \ \ if }\left\vert \phi
\right\vert >\beta_{\mathrm{MF}}.
\end{array}
\right.  \label{freeen}%
\end{equation}
where the dimensionless parameters, $\alpha_{MF}$ and $\beta_{MF}$, depend on
the interaction parameters as well as on the chemical potentials,
\begin{align}
\alpha_{\mathrm{MF}} &  =\frac{8}{3\pi}\left(  \frac{m_{F}}{m_{B}}\right)
\left[  a_{BB}k_{F}^{0}(\mu_{F})\right]  \left(  \frac{\mu_{F}}{\mu_{B}%
}\right)  ^{2},\\
\beta_{\mathrm{MF}}^{2} &  =\frac{\lambda_{BB}}{\lambda_{BF}}\frac{\mu_{F}%
}{\mu_{B}},
\end{align}
and we have introduced an effective Fermi momentum, $k_{F}^{0}(\mu_{F}%
)=\sqrt{(2m_{F}/\hbar^{2})\mu_{F}}$. In the density regime of interest
(enclosed by the I/II phase boundary and the line of spinodal decomposition),
the free energy ~Eq.\ (\ref{freeen}) generally exhibits $\phi$ dependency as
shown in Fig.~\ref{freeE}. This shape of the Landau free energy is associated
with a first-order phase transition: the system can reduce its free energy by
climbing or tunneling the free-energy barrier that separates two local minima.
In ordinary finite- temperature) first-order phase transitions, thermal
activation locally \textquotedblleft pushes\textquotedblright\ the system over
the barrier nucleating clumps of matter in the new phase. If the temperature is too low to initiate thermal activation, the
system can penetrate the barrier and nucleate by quantum-mechanical tunneling
\cite{Lifshitz}. The latter process, however, involves many-body tunneling,
which the tunneling rate generally scales exponentially with the number of
boson particles to be formed in each clump. A recent study \cite{Dima} found
that the tunneling rate exceeds the experimental lifetime of the system except
very close to the line of spinodal decomposition or for very strong
fermion-boson interactions. Otherwise, we can expect that at sufficiently low
temperatures and below the line of spinodal decomposition, the fermion-boson
mixture can remain in its homogeneous state for the duration of its
experimental lifetime even if that homogeneous state represents a metastable equilibrium.

We are also interested in what the mean-field approximation predicts for the
phase-separation dynamics in the spinodal decomposition region, $\rho_{F}%
^{0}>\rho_{F,\mathrm{crit}}$. We can understand the linear onset of the
instability dynamics from the study of the collective oscillations
\cite{HanPu,Yip}. Our studies \cite{Deborah1} revealed that two collective
modes can be excited: a zero-sound mode with its sound velocity modified by
boson-mediated fermion-fermion interactions, and a BEC sound mode modified by
the fermion mediated boson-boson interactions. As the system approaches the
line of spinodal decomposition, the fermion-mediated interactions reduce the
BEC sound velocity. This velocity vanishes at the line and becomes imaginary
above it, signaling the exponential growth of the long-wavelength modes. We
have estimated the time scale on which the corresponding fluctuations (which
drive the onset of the phase dynamics) grow \cite{Deborah2}. This analysis
suggests that the fermion-mediated interactions make the BEC unstable. The
vanishing of the sound velocity at the line of decomposition also suggests the
divergence of the boson compressibility, as the compressibility is inversely
proportional to the square of the sound velocity. However, the mean-field
equation of state does not exhibit this behavior. The mean field also fails to
include polaron effects: the mass of a fermion particle immersed in a
phonon-fluid can be significantly altered by its interactions with that fluid
\cite{Pines}. The importance of polaron effects was illustrated by a recent
study \cite{Fernando} that showed a single impurity (a distinguishable atom
immersed in a BEC) can self-localize into a small polaron-like state.

The rest of the paper deals with these problems by showing that the
correlation physics which we will include in the calculation of the equation
of state does lead to a divergence of the boson compressibility. The
perturbation treatment that we develop in this paper includes polaron effects.

\section{Linear-response description of fermion-mediated boson-boson
interactions\label{Sec_linearresponse}}

In this section, we derive the fermion-mediated interaction effects up to
lowest order by applying linear-response theory to a mean-field state. Writing
the energy response of the homogeneous many-body system as the small-amplitude
variation of the boson and fermion static densities, we obtain
\begin{align}
\rho_{B}\left(  \mathbf{r}\right)   &  =\rho_{B}^{0}+\delta\rho_{B}\left(
\mathbf{r}\right)  ,\label{rho B lr}\\
\rho_{F}\left(  \mathbf{r}\right)   &  =\rho_{F}^{0}+\delta\rho_{F}\left(
\mathbf{r}\right)  .\label{rho F lr}%
\end{align}
We also determine the corresponding variation in the boson-boson interaction
energy, $\delta E_{BB,\mathrm{int}}$, consistent with Eq. (\ref{MF1}). To
lowest order in the small-amplitude variation, the variation in interaction
energy takes the form
\begin{equation}
\delta E_{BB,\mathrm{int}}=\int d^{3}r\int d^{3}r^{\prime}\;\rho_{B}%
^{0}(\mathbf{r})v_{BB,\mathrm{tot}}\left(  \mathbf{r}-\mathbf{r}^{\prime
}\right)  \delta\rho_{B}\left(  \mathbf{r}^{\prime}\right)  ,\;\label{moden}%
\end{equation}
and from this equation we will determine the \textquotedblleft
total\textquotedblright\ interparticle interaction potential, $v_{\mathrm{tot}%
}\left(  \mathbf{r}-\mathbf{r}^{\prime}\right)  $, that includes the
fermion-mediated interaction potential, $v_{BB,\mathrm{fmed}}(\mathbf{r}%
-\mathbf{r}^{\prime})$, in addition to the usual boson-boson contact
interaction potential.

The boson density variation causes a modification of the overall mean-field
potential experienced by the fermion particles, $\delta V_{F}\left(
\mathbf{r}\right)  $, where%
\begin{equation}
\delta V_{F}\left(  \mathbf{r}\right)  =\lambda_{BF}\delta\rho_{B}\left(
\mathbf{r}\right)  .
\end{equation}
In linear-response theory, the fermion density variation $\delta\rho
_{F}\left(  \mathbf{r}\right)  $ caused by the static fermion potential
$\delta V_{F}\left(  \mathbf{r}\right)  $ is determined by the static response
function of the noninteracting fermions, $\chi_{F}^{0}$,
\begin{align}
\delta\rho_{F}\left(  \mathbf{r}\right)   &  =\int d^{3}r^{\prime}\chi_{F}%
^{0}\left(  \mathbf{r-r}^{\prime}\right)  \delta V_{F}\left(  \mathbf{r}%
^{\prime}\right)  \nonumber\\
&  =\lambda_{BF}\int d^{3}r^{\prime}\chi_{F}^{0}\left(  \mathbf{r-r}^{\prime
}\right)  \delta\rho_{B}\left(  \mathbf{r}^{\prime}\right)
.\label{delta rho_F}%
\end{align}
The above response function is the momentum Fourier transform of the dynamic
response function $\chi_{F}^{0}\left(  \mathbf{p},\omega\right)  $ in the
limit of vanishing frequency,
\begin{equation}
\chi_{F}^{0}\left(  \mathbf{r-r}^{\prime}\right)  =\int d^{3}p\;\chi_{F}%
^{0}\left(  \mathbf{p};\omega=0\right)  \exp\left[  i\mathbf{p}\cdot\left(
\mathbf{r-r}^{\prime}\right)  \right]  .
\end{equation}
In general, the dynamic density-density response function of a system confined
to a macroscopic volume $\Omega$ is
\begin{align}
\chi_{F}\left(  \mathbf{p},\omega\right)   &  =\frac{1}{\Omega}\sum
_{\left\vert \mathrm{int}\right\rangle }\left\vert _{F}\left\langle
\mathrm{int}\left\vert \hat{\rho}_{F}\left(  \mathbf{p}\right)  \right\vert
0\right\rangle _{F}\right\vert ^{2}\nonumber\\
&  \times\left(  \frac{1}{\hbar\omega-\left(  E_{\mathrm{int}}-E_{0}\right)
+i\eta}-\frac{1}{\hbar\omega+\left(  E_{\mathrm{int}}-E_{0}\right)  +i\eta
}\right)  ,\label{response}%
\end{align}
where $\hat{\rho}^{F}\left(  \mathbf{p}\right)  $ denotes the Fourier
transformed fermion density operator, $E_{0}$ represents the energy of the
unperturbed ground state $|0\rangle$, and the summation runs over all
intermediate excited fermion states $\left\vert \mathrm{int}\right\rangle $ of
excitation energy $E_{\mathrm{int}}-E_{0}$. At the end of the calculation, we
take the limit $\eta\rightarrow0$ while approaching zero from the positive
side $\eta>0$. In the static limit of a noninteracting fermion system for
which the unperturbed ground state corresponds to a filled Fermi sphere and
for which the intermediate states are particle-hole excitations, $\chi_{F}%
^{0}\left(  \mathbf{p};\omega=0\right)  $ yields
\begin{equation}
\chi_{F}^{0}\left(  \mathbf{p};\omega=0\right)  =-\frac{2}{\Omega}%
\sum_{\left\vert \mathrm{int}\right\rangle }\frac{n_{\mathbf{k}}^{F}\left(
1-n_{\mathbf{k}+\mathbf{p}}^{F}\right)  }{\epsilon_{\mathbf{k}+\mathbf{p}}%
^{F}-\epsilon_{\mathbf{k}}^{F}}\;,\label{fermresp}%
\end{equation}
where $\epsilon_{\mathbf{k}}^{F}$ is the excitation energy of a single fermion
of momentum $\mathbf{k}$, and where $n_{\mathbf{k}}^{F}$ denotes the
zero-temperature occupation number of the single-particle $\mathbf{k}$
momentum state. Then, the interaction energy caused by $\delta\rho_{B}$ and
$\delta\rho_{F}$ is equal to
\begin{align}
\delta E_{\mathrm{int}} &  =\int d^{3}r\;\rho_{B}^{0}\left(  \mathbf{r}%
\right)  \lambda_{BB}\delta\rho_{B}\left(  \mathbf{r}\right)  \nonumber\\
&  +\int d^{3}r\text{ }\rho_{B}^{0}\left(  \mathbf{r}\right)  \lambda
_{BF}\delta\rho_{F}\left(  \mathbf{r}\right)  \nonumber\\
&  +\int d^{3}r\;\rho_{F}^{0}\left(  \mathbf{r}\right)  \lambda_{BF}\delta
\rho_{B}\left(  \mathbf{r}\right)  .\label{deint}%
\end{align}
Inserting Eq.\ (\ref{delta rho_F}) into the mean-field energy of
Eq.\ (\ref{deint}), we obtain the corresponding modification of the
interaction energy,
\begin{align}
\delta E_{\mathrm{int}} &  =\int d^{3}r\;\rho_{B}^{0}\left(  \mathbf{r}%
\right)  \lambda_{BB}\delta\rho_{B}\left(  \mathbf{r}\right)  \nonumber\\
&  +\int d^{3}r\;\rho_{F}^{0}\left(  \mathbf{r}\right)  \lambda_{BF}\delta
\rho_{B}\left(  \mathbf{r}\right)  \nonumber\\
&  +\int d^{3}r\int d^{3}r^{\prime}\rho_{B}^{0}\left(  \mathbf{r}\right)
\lambda_{BF}^{2}\chi_{F}^{0}\left(  \mathbf{r-r}^{\prime}\right)  \delta
\rho_{B}\left(  \mathbf{r}^{\prime}\right)  .\label{delta E_int}%
\end{align}
A comparison of the first and third terms of the right-hand side of
Eq.~(\ref{delta E_int}) to Eq.~(\ref{moden}) indicates that we can associate
\begin{equation}
v_{BB,\mathrm{tot}}\left(  \mathbf{r-r}^{\prime}\right)  =\lambda_{BB}%
\delta\left(  \mathbf{r-r}^{\prime}\right)  +v_{BB,\mathrm{fmed}}\left(
\mathbf{r-r}^{\prime}\right)
\end{equation}
with the linear-response fermion-mediated interaction equal to
\begin{equation}
v_{BB,\mathrm{fmed}}\left(  \mathbf{r-r}^{\prime}\right)  =\lambda_{BF}%
^{2}\chi_{F}^{0}\left(  \mathbf{r-r}^{\prime}\right)  .
\end{equation}

To determine the fermion-mediated boson-boson energy contribution to the
equation of state of the fermion-BEC mixtures, we integrate the
fermion-mediated boson-boson interaction potential over the boson density,
\begin{align}
E_{BB,\mathrm{fmed}} &  \approx\int d^{3}r\int d^{3}r^{\prime}\;\rho_{B}%
^{0}\left(  \mathbf{r}\right)  \frac{v_{BB,\mathrm{fmed}}\left(
\mathbf{r-r}^{\prime}\right)  }{2}\rho_{B}^{0}\left(  \mathbf{r}^{\prime
}\right)  \nonumber\\
&  =\frac{\lambda_{BF}^{2}}{2}\chi_{F}^{0}\left(  \mathbf{p}\rightarrow
0;\omega=0\right)  \frac{N_{B}^{2}}{\Omega}\;\;,\label{meden}%
\end{align}
a result that we will obtain rigorously from a perturbation treatment.

While we can evaluate $\chi_{F}^{0}\left(  \mathbf{p}\rightarrow
0;\omega=0\right)  $ directly from Eq.~(\ref{fermresp}), we can also derive it
from the Thomas-Fermi approximation as the static ($\omega\rightarrow0$),
long-wavelength ($\left\vert \mathbf{k}\right\vert \ll k_{F}$) limit coincides
with the regime of validity of the Thomas-Fermi description. In this
approximation, we introduce a local Fermi momentum $k_{F}\left(
\mathbf{r}\right)  $ whose local Fermi energy equals the difference of the
system's chemical potential and effective Fermion potential energy,
\begin{equation}
\frac{\hbar^{2}\left[  k_{F}\left(  \mathbf{r}\right)  \right]  ^{2}}{2m_{F}%
}=\mu_{F}-\lambda_{BF}\rho_{B}^{0}-\lambda_{BF}\delta\rho_{B}\left(
\mathbf{r}\right)  .
\end{equation}
Relative to the equilibrium momentum $k_{F}^{0}$, and its corresponding Fermi
energy $\hbar^{2}\left(  {k_{F}^{0}}\right)  ^{2}/2m_{F}=\mu_{F}-\lambda
_{BF}\rho_{B}^{0}$, the Fermi-momentum variation $\delta k_{F}\left(
\mathbf{r}\right)  $, where $k_{F}=k_{F}^{0}+\delta k_{F}$, becomes
\begin{equation}
\delta k_{F}\left(  \mathbf{r}\right)  \simeq-\frac{m_{F}}{\hbar^{2}k_{F}^{0}%
}\lambda_{BF}\delta\rho_{B}\left(  \mathbf{r}\right)  .
\end{equation}
Using $\rho_{F}=k_{F}^{3}/\left(  6\pi^{2}\right)  $, the corresponding linear
variation of the fermion density $\delta\rho_{F}\left(  \mathbf{r}\right)  $
is
\begin{equation}
\delta\rho_{F}\left(  \mathbf{r}\right)  \simeq\frac{\left(  k_{F}^{0}\right)
^{2}}{2\pi^{2}}\delta k_{F}\left(  \mathbf{r}\right)  =-\frac{m_{F}k_{F}^{0}%
}{\hbar^{2}2\pi^{2}}\lambda_{BF}\delta\rho_{B}\left(  \mathbf{r}\right)  ,
\end{equation}
from which we can extract the long wavelength static density-density response
function
\begin{equation}
\lim_{k\rightarrow0}\chi_{F}^{0}\left(  \mathbf{k},\omega=0\right)
=-\frac{m_{F}k_{F}^{0}}{\hbar^{2}2\pi^{2}}.
\end{equation}
With this expression and from Eq.~(\ref{meden}), we obtain the fermion
mediated boson-boson interaction energy contribution to the mixture's equation
of state,
\begin{equation}
E_{BB,\mathrm{fmed}}=-\lambda_{BF}\frac{N_{B}^{2}}{2\Omega}\left(
1+\frac{m_{F}}{m_{B}}\right)  \frac{a_{BF}k_{F}^{0}}{\pi},\;\label{EBBfmed}%
\end{equation}
which gives the fermion-mediated boson-boson interaction energy contribution
that is consistent with linear response.

We note that the fermion-mediated energy in Eq.\ (\ref{EBBfmed}) not only
arises in response to a boson density variation but also is an integral part
of the many-body energy, as we will show in this section.

\section{Fermion mediated boson-boson interactions in second-order
perturbation\label{Sec_FmedBB}}

In this section, we determine the fermion-mediated boson-boson interactions
using a perturbation approach.

Two assumptions were made. One is that the mixture is homogeneous, i.e., the
average densities $\rho_{B}^{0}$ and $\rho_{F}^{0}$ are position independent,
and the other is that the boson-boson scattering length, $a_{BB}$, is
sufficiently small to ensure that the gas parameter $\sqrt{\rho_{B}^{0}%
a_{BB}^{3}}\ll1$ so that the BEC is dilute and well-described by the
Bogoliubov approximation. The zeroth order corresponds to the limit
$\lambda_{BF}\rightarrow0$ while $\lambda_{BB}$ remains constant. Thus, the
zeroth-order ground state is a product state $\left\vert 0\right\rangle
=\left\vert 0\right\rangle _{B}\otimes\left\vert 0\right\rangle _{F}$, where
$\left\vert 0\right\rangle _{B}$ denotes the weakly interacting BEC ground
state, and $\left\vert 0\right\rangle _{F}$ denotes the ground state of an
ideal gas of single spin component indistinguishable fermions, corresponding
to a filled Fermi sphere of radius $k_{F}$. Similarly, the excited states of
the zeroth-order Hamiltonian are product states of the zeroth order fermion
ground-state and the BEC ground state (or alternatively particle-hole states
and boson quasiparticle states).

\subsection{The perturbation treatment}

The perturbation Hamiltonian, $\hat{H}_{p}$, is the effective interaction
potential that describes the short-range boson-fermion interactions,
\begin{equation}
\hat{H}_{p}=\frac{\lambda_{BF}}{\Omega}\sum_{\mathbf{k}}\hat{\rho}%
_{\mathbf{k}}^{B}\hat{\rho}_{-\mathbf{k}}^{F}, \label{hp}%
\end{equation}
where $\hat{\rho}_{\mathbf{k}}^{B}$ and $\hat{\rho}_{{-\mathbf{k}}}^{F}$ are
the momentum Fourier transform of the boson and fermion density operators, respectively.

Using the Bogoliubov approximation for weakly interacting bosons, we write
\begin{equation}
\hat{\rho}_{\mathbf{k}}^{B}\approx N_{B}\left(  \hat{c}_{B,\mathbf{k}%
}^{\dagger}+\hat{c}_{B,-\mathbf{k}}\right)  ,
\end{equation}
where $\hat{c}^{\dagger}$ $\left(  \hat{c}\right)  $ is the boson particle
creation (annihilation) operator that is Bogoliubov transformed into
quasi-particle (phonon) operators, $\eta^{\dagger}$ and $\eta$, where
$\eta|0\rangle_{B}=0$. The Bogoliubov transformation also yields the energy
cost of exciting a single quantum of the collective BEC oscillation given by
the Bogoliubov dispersion $\hbar\omega_{k}^{B}=\hbar kc\sqrt{1+\left(
k\xi\right)  ^{2}}$, where $c$ denotes the sound velocity of the unperturbed
BEC, $c=\sqrt{\lambda_{BB}\rho_{B}^{0}/m_{B}}$ and $\xi$ represents its
coherence length $\xi=\left(  16\pi\rho_{B}^{0}a_{BB}\right)  ^{-1/2}$. A
direct application of the Bogoliubov transformation also gives the excitation
density matrix element,
\begin{equation}
_{B}\left\langle \mathbf{k}\left\vert \hat{\rho}_{\mathbf{k}}^{B}\right\vert
0\right\rangle _{B}=\sqrt{\frac{\hbar^{2}k^{2}/2m_{B}}{\hbar\omega_{k}^{B}}%
}\sqrt{N_{B}}.\label{bd}%
\end{equation}
The many-body energy can be calculated by expanding the term in the
fermion-boson interaction strength,
\begin{equation}
\Delta E=E-E_{0}=\Delta E_{1}+\Delta E_{2}+\cdots\,,
\end{equation}
where the $j$th-order contribution, $\Delta E_{j}$, varies as $\sim\left(
\lambda_{BF}\right)  ^{j}$ as the boson-fermion scattering length is modified.

The first-order energy contribution gives the mean-field fermion-boson
interaction energy,
\begin{equation}
\Delta E_{1}=_{B}\langle0|_{F}\langle0|\hat{H}_{p}|0\rangle_{F}|0\rangle
_{B}=\lambda_{BF}\frac{N_{B}N_{F}}{\Omega}.
\end{equation}
Hence, since we can neglect the depletion contribution to the many-body ground
state of the zeroth-order BEC (which gives a relative error of order
$\sim\sqrt{\rho_{B}^{0}a_{BB}^{3}}$), the zeroth-order and first order energy
terms give an equation of state that is identical to the mean-field equation
of state. Thus, the perturbation terms of higher order add correlation terms
to the mean-field equation of state.

The second order contribution can be written as the sum over all intermediate
states, $|\mathrm{int}\rangle$,
\begin{equation}
\Delta E_{2}=-\sum_{|\mathrm{int}\rangle}\frac{|\langle\mathrm{int}|\hat
{H}_{p}|0\rangle|^{2}}{E_{\mathrm{int}}-E_{0}},
\end{equation}
where $E_{\mathrm{int}}$ denotes the corresponding (zeroth-order) excitation
energy of the intermediate state. Here we distinguish three kinds of
intermediate states: The first category consists of products of the fermion
ground state and quasiparticle boson excited states. The second category
comprises products of the BEC ground state with particle-hole fermion excited
states. The third category consists of products of excited fermion and excited
boson states. It is the second category that yields the fermion-mediated
boson-boson interaction energy, $\Delta E_{BB,\mathrm{fmed}}^{\left(
2\right)  }$,
\begin{equation}
\Delta E_{BB,\mathrm{fmed}}^{\left(  2\right)  }=-\left(  \frac{\lambda_{FB}%
}{\Omega}\right)  ^{2}\sum_{\mathbf{k},\mathbf{p}}\left\vert _{B}\langle
0|\hat{\rho}_{\mathbf{k}}^{B}|0\rangle_{B}\right\vert ^{2}\frac{n_{\mathbf{p}%
}^{F}\left(  1-n_{\mathbf{p}+\mathbf{k}}^{F}\right)  }{\epsilon_{\mathbf{p}%
+\mathbf{k}}^{F}-\epsilon_{\mathbf{p}}^{F}},\label{fmed2}%
\end{equation}
equal to $E_{BB,\mathrm{fmed}}$ in Eq.\ (\ref{meden}). In the limit where the
system size becomes infinite, the matrix element tends to $_{B}\langle
0|\hat{\rho}_{\mathbf{k}}^{B}|0\rangle_{B}\rightarrow N_{B}\delta_{\mathbf{k}%
}$, implying that the corresponding process involves zero momentum transfer.
Perhaps because of the zero momentum nature of the momentum transfer
processes, this contribution has been left out in other papers
\cite{Viverit2,Wilkens02,Wilkens03}. Here, we simply mention that this term
needs to be included and we show explicitly how the long-wavelength limit
corresponds to the infinite size limit of the fermion system.

Consider a large (i.e., a linear size that significantly exceeds the BEC
coherence length) but finite-size BEC system immersed in a homogeneous
infinite Fermi sea. Then, let the BEC size approach that of the homogeneous
fermion system. A finite BEC has a density expectation value $_{F}%
\langle0|\hat{\rho}_{\mathbf{k}}^{B}|0\rangle_{B}$ that is a smooth function
of $\mathbf{k}$. A Fourier transform of the average spatial distribution
gives
\begin{equation}
_{F}\langle0|\hat{\rho}_{\mathbf{k}}^{B}|0\rangle_{B}=\int d^{3}r\;\exp\left(
i\mathbf{k}\cdot\mathbf{r}\right)  \;\rho_{B}^{0}(\mathbf{r}).\label{ft}%
\end{equation}
Inserting the expression of Eq.~(\ref{ft}) into Eq.~(\ref{fmed2}), we obtain
\begin{align}
\Delta E_{BB,\mathrm{fmed}}^{\left(  2\right)  } &  =\int d^{3}r\;\int
d^{3}r^{\prime}\;\rho_{B}^{0}(\mathbf{r})\left[  -\lambda_{BF}^{2}\int
\frac{d^{3}k}{\left(  2\pi\right)  ^{3}}\exp\left[  i\mathbf{k}\cdot\left(
\mathbf{r}-\mathbf{r}^{\prime}\right)  \right]  \frac{1}{\Omega}%
\sum_{\mathbf{p}}\frac{n_{\mathbf{p}}^{F}\left(  1-n_{\mathbf{p}+\mathbf{k}%
}^{F}\right)  }{\epsilon_{\mathbf{p}+\mathbf{k}}^{F}-\epsilon_{\mathbf{k}}%
^{F}}\right]  \rho_{B}^{0}(\mathbf{r}^{\prime})\;\nonumber\\
&  =\int d^{3}r\;\int d^{3}r^{\prime}\;\rho_{B}^{0}(\mathbf{r})\frac
{V_{BB,\mathrm{fmed}}^{\left(  0\right)  }\left(  \mathbf{r}-\mathbf{r^{\prime
}}\right)  }{2}\rho_{B}\left(  \mathbf{r}^{\prime}\right)  .
\end{align}
This expression is independent of the BEC size and should clearly be included
in the many-body energy regardless of the size of the BEC. The infinite BEC
limit simplifies the expression by allowing a straightforward substitution to
center-of-mass and relative coordinates, equivalent to the long-wavelength
limit
\begin{equation}
\left\langle E_{BB,\mathrm{fmed}}^{\left(  0\right)  }\right\rangle
=-\frac{N_{B}^{2}}{2\Omega}\lim_{\mathbf{k}\rightarrow0}v_{BB,\mathrm{fmed}%
}^{\left(  0\right)  }(\mathbf{k}),
\end{equation}
where $v_{BB,\mathrm{fmed}}^{\left(  0\right)  }(\mathbf{k})$ is the Fourier
transform of $v_{BB,\mathrm{fmed}}(\mathbf{x})$ in Eq.\ (\ref{meden}).

It is interesting to note that by adding the fermion-mediated boson-boson
energy to the mean-field ground-state energy Eq.\ (\ref{MFEOS}), one obtains
an equation of state that exhibits an instability at the same fermion density
$\rho_{F,\mathrm{crit}}$ as the mean-field equation of state. However, the
instability is not a saddlepoint instability but one of diverging boson
compressibility. The boson chemical potential derivative is equal to
\begin{align}
\frac{\partial\mu_{B}}{\partial\rho_{B}} &  =\lambda_{BB}\left[  1-\left(
\frac{\lambda_{BF}}{\lambda_{BB}}\right)  \left(  1+\frac{m_{F}}{m_{B}%
}\right)  \frac{k_{F}a_{FB}}{\pi}\right]  \nonumber\\
&  =\lambda_{BB}\left[  1-\left(  1+\frac{m_{F}}{m_{B}}\right)  \left(
1+\frac{m_{B}}{m_{F}}\right)  \frac{a_{FB}}{a_{BB}}\frac{a_{BF}k_{F}}{2\pi
}\right]  ,
\end{align}
which vanishes as $\rho_{F}\rightarrow\rho_{F,\mathrm{crit}}$. Since the boson
compressibility $\kappa_{B}=\left[  \left(  \rho_{B}^{0}\right)  ^{2}%
\partial\mu_{B}/\partial\rho_{B}\right]  ^{-1}$, the compressibility diverges
at that same fermion density. That large values of the compressibility can
make a significant impact on the physical behavior of systems is illustrated
by the phenomenon of critical opalescence--the sudden increase in light
scattering that can turn the appearance of systems milky white near the
critical point--which is caused by the increase of density fluctuations that
accompanies the large value of the compressibility. In trapped systems, we
expect the increase of compressibility to result in higher densities and
smaller BEC size. The strikingly different density profile could be an easily
detectable \textquotedblleft beyond-mean-field\textquotedblright\ effect.

\section{Linked cluster expansion of the ground-state energy of fermion-boson
mixtures\label{Sec_cluster}}

\begin{figure}[ptb]
\begin{center}
\includegraphics[height=2in, width=3in] {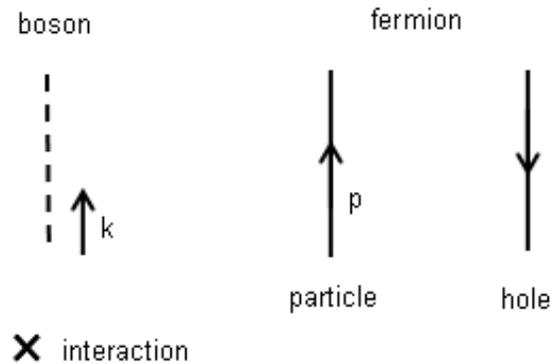}
\end{center}
\caption{Symbols and directions used for the diagrams in this paper.}%
\label{symbol}%
\end{figure}

In this section, we develop a linked cluster expansion that complements and
extends the above second-order result. The starting point of the expansion is
Goldstone's theorem \cite{Gold}, which states that the difference of the exact
many-body energy, $E$, and the zeroth-order energy, $E_{0}$, can be written as
an infinite sum of zeroth-order ground-state expectation values,
\begin{equation}
E-E_{0}=\sum_{n}\left\langle 0\left\vert \left(  \hat{H}_{p}\frac{1}{\hat
{H}_{0}-E_{0}}\right)  ^{n}\hat{H}_{p}\right\vert 0\right\rangle
_{L},\label{lc}%
\end{equation}
where the subscript $L$ indicates that only \textquotedblleft
linked\textquotedblright\ diagrams are included (unlinked diagrams factor out
to cancel the normalization factor in the denominator). The expectation values
of the perturbation Hamiltonian in Eq.~(\ref{lc}) are obtained by inserting
the completeness relation as a sum over zeroth-order eigenstates in between
each perturbation operator, $\hat{H}_{p}$, and propagator operator, $\left(
\hat{H}_{0}-E_{0}\right)  ^{-1}$. Since the bra and ket state matrix elements
of the propagator are eigenstates of $\hat{H}_{0}$, the matrix elements are
diagonal and give rise to unperturbed energy denominators. The interaction
elements in the numerators (the matrix elements of $\hat{H}_{p}$ ) are
nondiagonal but are easily calculated. We envision the order in which the
operator matrix elements occur to correspond to an effective time ordering
(right to left corresponds to increasing \textquotedblleft
time\textquotedblright). We represent each term of the series of interaction
matrix elements and energy denominators (propagators) as a diagram where the
effective time parameter runs upwards (see Fig.\ \ref{symbol}).

\begin{figure}[ptb]
\begin{center}
\includegraphics[height=2.4512in, width=3.3996in] {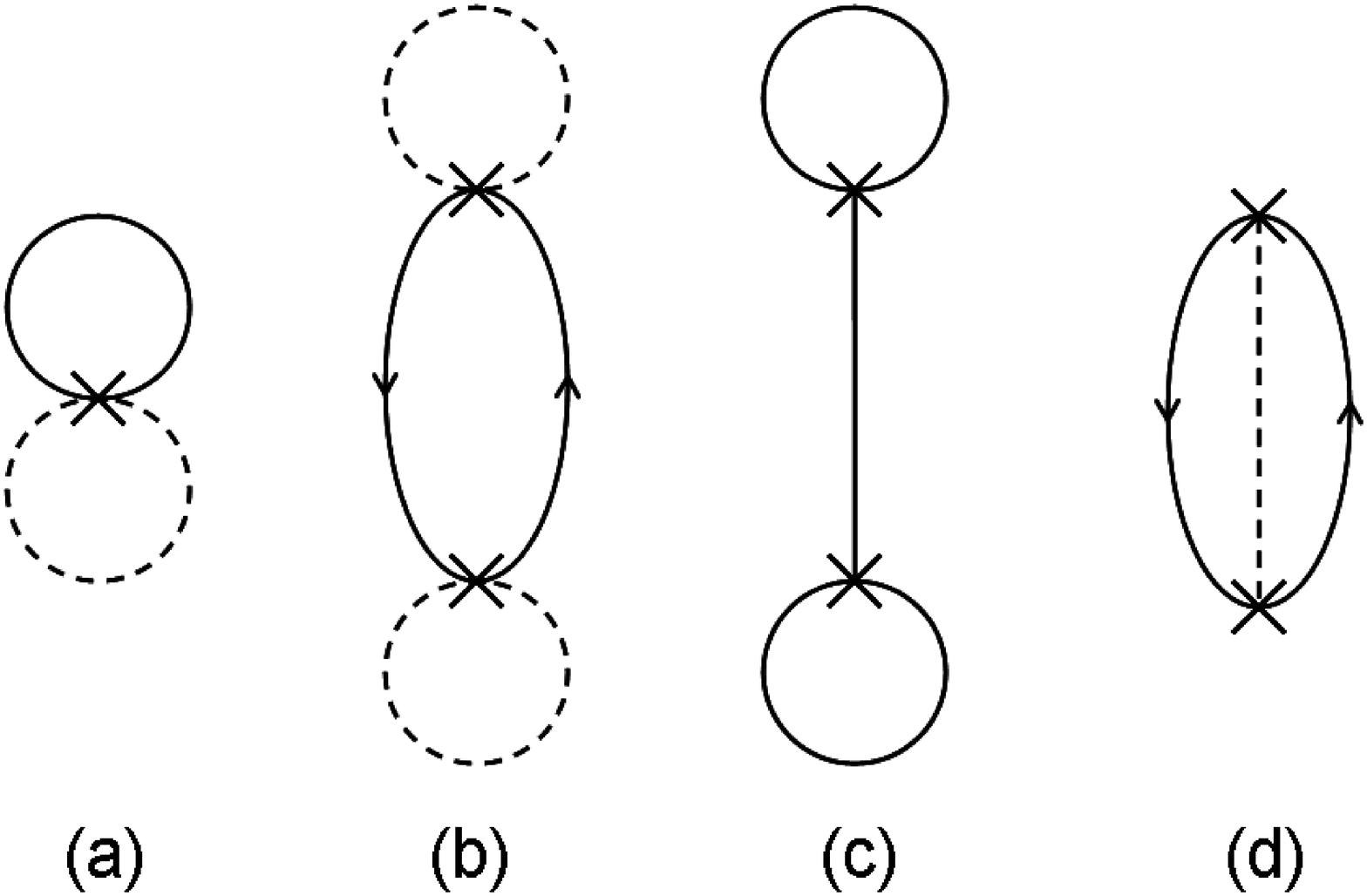}
\end{center}
\caption{First- and second-order Goldstone diagrams for a single spin
component fermion-boson BEC mixture. (a) Contact interaction, (b)
fermion-mediated boson-boson interaction, (c) boson-mediated fermion-fermion
interaction (direct term), and (d) boson-mediated fermion-fermion interaction
(exchange and polaron terms).}%
\label{diagram}%
\end{figure}

In Fig.\ \ref{diagram}, a \textquotedblleft cross\textquotedblright, $\times$,
denotes a vertex that gives rise to an interaction matrix element. The crosses
are connected by fermion and/or boson lines that represent the intermediate
states that propagate from one interaction matrix element to another and give
rise to an energy denominator. The excited fermion states are represented by
solid lines with arrows: upward arrows indicate \textquotedblleft
particle\textquotedblright\ states and downward arrows \textquotedblleft
hole\textquotedblright\ states. The boson quasiparticle states are represented
by dashed lines. Although we do not show the phonon time arrows explicitly in
our diagrams, we treat the dashed phonon lines as upward-pointing (boson
excitations do not have holes associated with them) when imposing momentum
conservation at each vertex (see below). Then the propagator energy
denominators are sums over particle-hole fermion excitation energies and boson
quasiparticle energies.

Since each application of the $\hat{H}_{p}$ operator of Eq.~(\ref{hp}) creates
and annihilates a fermion particle and either creates or annihilates a boson
quasiparticle (or gives rise to the diagonal ground-state boson density matrix
element), each $\times$ vertex has to have a fermion line that arrives and one
that leaves, and each vertex has to have a boson line that either arrives or
leaves. The exception to vertices that have a boson line either leaving or
arriving are those vertices in which a boson line curves back to make a full
circle--a 'loop'--which corresponds to the diagonal ground-state density
expectation value $_{B}\left\langle 0\left\vert \hat{\rho}^{B}\right\vert
0\right\rangle _{B}$ that occurs in the long-wavelength limit arising in the
second-order evaluation of the fermion-mediated interaction discussed above.
Similarly, a full line that loops back upon the $\times$ vertex that it left
from indicates a fermion ground-state matrix element, $_{F}\left\langle
0\left\vert \hat{\rho}_{\mathbf{k}}^{F}\right\vert 0\right\rangle
_{F}\rightarrow\delta_{\mathbf{k}}N_{F}$.

The rules for translating each diagram to a numerical factor are
straightforward, so we briefly describe the procedure here. In our case of a
homogeneous system, the particle and quasiparticle excitations correspond to
good momentum quantum numbers and we assign momentum labels $\mathbf{p}%
_{1},\mathbf{p}_{2},...$ to the fermion lines and $\mathbf{k}_{1}%
,\mathbf{k}_{2},...$ to the boson lines, ensuring conservation of momentum at
each $\times$. Next, we divide the vertical `time' axis into separate time
intervals between the different interaction $\times$ points and determine the
propagator energy denominators to multiply. Converting the remaining sums over
boson $\mathbf{k}_{j}$ and fermion $\mathbf{p}_{i}$ momenta to dimensionless
integrals introduces dimensionless expansion parameters $\alpha_{F}$ and
$\alpha_{B}$ that quantify the relative importance of fermion and boson
excitation induced correlations. Then, we assign a single factor $\left\vert
\lambda_{BF}\right\vert /\Omega$ of the $\hat{H}_{p}$ operator to each of the
momentum summations that remain after enforcing conservation of momentum, and
multiply and divide by $\mu_{B}^{0}$ for the boson sums and by $\mu_{F}^{0}$
for the fermion momentum sums.

After taking the infinite system limit $\Omega^{-1}\sum_{\mathbf{k}%
}\rightarrow\left(  2\pi\right)  ^{-3}\int d^{3}k$, we introduce dimensionless
momentum variables $\mathbf{k}_{j}^{\prime}=\xi\mathbf{k}_{j}$ for bosons,
where $\xi=\left(  16\pi\rho_{B}^{0}a_{BB}\right)  ^{-1/2}$ is the BEC
coherence length, and $\mathbf{p}_{i}^{\prime}=\mathbf{p}_{i}/k_{F}$, where
$k_{F}$ is the Fermi momentum, and $\hbar^{2}k_{F}^{2}/\left(  2m_{F}\right)
=\mu_{F}^{0}$. Then integrate over fermion momenta, and write the zeroth-order
fermion and boson chemical potentials as $\mu_{F}^{0}$ and $\mu_{B}^{0}$, respectively.

For boson momentum integrals we obtain
\begin{align}
\frac{1}{\mu_{B}^{0}}\frac{\left\vert \lambda_{BF}\right\vert }{\Omega}%
\sum_{\mathbf{k}} &  \rightarrow\int d^{3}k\frac{1}{\mu_{B}^{0}}%
\frac{\left\vert \lambda_{BF}\right\vert }{\left(  2\pi\right)  ^{3}%
}\nonumber\\
&  =\left(  \frac{\left\vert \lambda_{BF}\right\vert \xi^{-3}}{\mu_{B}%
^{0}\left(  2\pi\right)  ^{2}}\right)  \frac{1}{2\pi}\int d^{3}k^{\prime
}\nonumber\\
&  =\alpha_{B}\frac{1}{2\pi}\int d^{3}k^{\prime},
\end{align}
where $\alpha_{B}$ denotes the boson expansion parameter
\begin{equation}
\alpha_{B}=\left(  1+\frac{m_{B}}{m_{F}}\right)  \frac{2}{\pi}\frac{\left\vert
\lambda_{BF}\right\vert }{\xi},
\end{equation}
and for the fermion momentum sum,
\begin{equation}
\frac{1}{\mu_{F}^{0}}\frac{|\lambda_{BF}|}{\Omega}\sum_{\mathbf{p}}=\alpha
_{F}\frac{1}{2\pi}\int d^{3}p^{\prime},
\end{equation}
where
\begin{equation}
\alpha_{F}=\frac{1}{\mu_{F}^{0}}\frac{|\lambda_{BF}|k_{F}^{3}}{4\pi^{2}%
}=\left(  1+\frac{m_{F}}{m_{B}}\right)  \frac{|a_{FB}|k_{F}}{\pi},
\end{equation}
which plays the role of fermion expansion parameter.

The energy denominators obtained in this procedure contain fermion and boson
excitation energies. So, first, we make the denominators positive valued,
introducing a $-1$ factor for each propagator, and verify that particle
excitation energies appear with a positive sign and hole excitations with a
negative sign. Then, we make the propagators dimensionless by bringing out an
energy scaling factor, $\mu_{F}^{0}$ or $\mu_{B}^{0}$,
\begin{align}
\frac{1}{\epsilon_{\mathbf{p}_{1}+\mathbf{p}_{2}}^{F}-\epsilon_{\mathbf{p}%
_{2}}^{F}} &  =\frac{1}{\mu_{F}^{0}}\frac{1}{\mathbf{p}_{1}^{\prime}%
\cdot\mathbf{p}_{1}+2\mathbf{p}_{1}^{\prime}\cdot\mathbf{p}_{2}^{\prime}%
},\nonumber\\
\frac{1}{\epsilon_{\mathbf{k}}} &  =\frac{1}{2\mu_{B}^{0}}\frac{1}{k^{\prime
}\sqrt{1+k^{\prime}{}^{2}}}.
\end{align}
For those propagators that contain both boson and fermion excitation energies,
we have to choose which energy scale to use. Scaling by the boson chemical
potential, we obtain the following expression for the propagator that
corresponds to the excited state with one BEC phonon excited of momentum
$-\mathbf{k}$, a fermion particle excitation of momentum $\mathbf{p}%
+\mathbf{k}$, and a fermion hole excitation of momentum $\mathbf{p}$,
\begin{equation}
\frac{1}{\epsilon_{-\mathbf{k}}^{B}+\epsilon_{\mathbf{p}+\mathbf{k}}%
^{F}-\epsilon_{\mathbf{p}}^{F}}=\frac{1}{2\mu_{B}^{0}}\frac{1}{k^{\prime}%
\sqrt{1+k^{\prime}{}^{2}}+\left(  m_{B}/m_{F}\right)  \left(  \mathbf{k}%
^{\prime}\cdot\mathbf{k}^{\prime}+2\alpha\mathbf{p}^{\prime}\cdot
\mathbf{k}^{\prime}\right)  },
\end{equation}
where $\mathbf{k}^{\prime}\equiv\xi\mathbf{k}$, $\mathbf{p}^{\prime}%
\equiv\mathbf{p}/k_{F}$, and $\alpha\equiv\xi k_{F}$. Hence, the mixed
fermion-boson propagators give rise to dimensionless integrands that depend on
the mass ratio $\left(  m_{F}/m_{B}\right)  $ as well as on the $\alpha$
parameter that compares the BEC and fermion lengthy scales: $\alpha=2\left(
m_{B}/m_{F}\right)  \left(  \alpha_{F}/\alpha_{B}\right)  $.

We now include the contributions that stem from the interaction matrix
elements. Each fermion loop contributes a factor $N_{F}$, each boson loop
contributes $N_{B}$, and each boson line that connects two vertices and
carries a momentum $\mathbf{k}_{j}$ gives rise to a factor
\begin{equation}
N_{B}\frac{\hbar^{2}k_{j}^{2}/\left(  2m_{B}\right)  }{\epsilon_{\mathbf{k}%
_{j}}^{B}}=N_{B}\left(  \frac{k_{j}^{\prime}}{\sqrt{1+k_{j}^{\prime}{}^{2}}%
}\right)  .
\end{equation}
Each fermion hole line of momentum $\mathbf{p}_{i}$ gives a factor
$n_{\mathbf{p}_{i}}^{F}$, where $n^{F}$ denotes the zero-temperature
Fermi-Dirac distribution function. In scaled momentum units,
\begin{equation}
n_{\mathbf{p}_{i}}^{F}\rightarrow n_{\mathbf{p}_{i}^{\prime}}^{F}%
=\theta\left(  1-\left\vert \mathbf{p}_{i}^{\prime}\right\vert \right)  ,
\end{equation}
where $\theta$ represents the usual Heaviside function: $\theta\left(
x\right)  =1$ if $x>0$ and $\theta\left(  x\right)  =0$ if $x<0$. Likewise,
each upward fermion line of momentum $\mathbf{p}_{k}$ gives the factor
\begin{equation}
1-n_{\mathbf{p}_{k}}^{F}\rightarrow1-n_{\mathbf{p}_{k}^{\prime}}^{F}%
=\theta\left(  \left\vert \mathbf{p}_{k}^{\prime}\right\vert -1\right)  ,
\end{equation}
associated with a particle fermion excitation.

Finally, we have to specify how to take the infinite system (long-wavelength)
limit associated with those diagrams that have loops. From the calculation of
the fermion-mediated boson-boson interaction in the preceding section, we
deduce the following rule: We assign a fictitious momentum $\mathbf{p}_{L}$
with each fermion loop and a momentum $\mathbf{k}_{L}$ for the boson loops,
conserve momentum at each vertex with this additional momentum, and then take
the long-wavelength limits $\lim k_{L}\rightarrow0$, $\lim p_{L}\rightarrow0$
at the end of the calculation.

In the next two sections, we use the procedure described above for calculating
the equation of state and then analyze the results.

\section{Low-order diagrammatic analysis\label{Sec_low_order}}

In this section, we briefly describe the first- and second-order contributions
to the many-particle ground-state energy of a homogeneous fermion-BEC mixture.
Below, we describe the numerical contributions of each diagram shown in
Figs.\ \ref{diagram}$\left(  b\right)  $-\ref{diagram}$\left(  d\right)  $.
The derivation of the full expressions is beyond the scope of this paper and
will be reported elsewhere. The analysis of this section serves to reveal
trends, such as the vanishing of specific diagrams in the low-density BEC
limit $\alpha_{B}\rightarrow0$ in which BEC fluctuation-induced correlations
can be neglected.

Much of the challenge in calculating the numerical contributions goes into
making judicious choices in dealing with the different ways that the rules can
be applied. The labeling of the fermion and boson lines with momenta can be
carried out in different ways. For instance, in calculating the contribution
of diagram $\left(  c\right)  $, \textquotedblleft the bubble diagram with
intersecting phonon line\textquotedblright, one can either assign a boson
momentum $\mathbf{k}$ to the phonon line and a momentum $\mathbf{p}%
-\mathbf{k}$ to the fermion particle line or one can assign a momentum
$\mathbf{p}^{\prime}$ to the fermion particle line so that the phonon line
acquires a momentum $\mathbf{p}-\mathbf{p}^{\prime}$. The first choice results
in an integral over a fermion momentum and an integral over a boson momentum,
whereas the second choice results in a double fermion momentum integral. Also,
one can scale the mixed fermion-boson energy denominators in the propagators
either by the zeroth-order fermion or by the zeroth-order boson chemical
potential. Neither of these choices affects the final result, but they can
obscure common factors and common limits of different diagrams. For instance,
diagrams $\left(  b\right)  $ and $\left(  c\right)  $ give contributions that
are equal in magnitude and opposite in sign in the limit $\alpha\rightarrow0$.

\begin{figure}[ptb]
\begin{center}
\includegraphics[height=2.5555in, width=3.4653in] {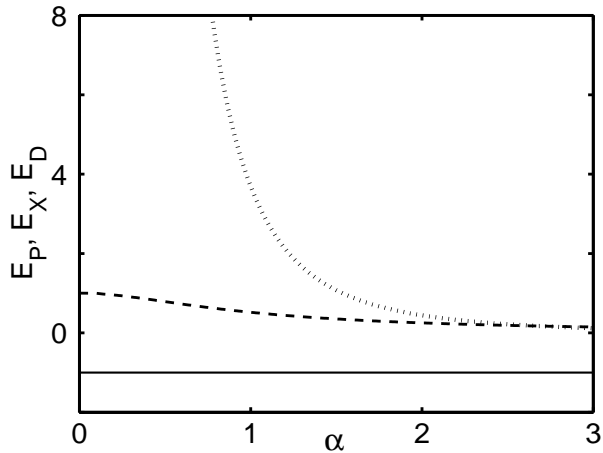}
\end{center}
\caption{Direct $E_{D}$, exchange $E_{X}$, and polaron $E_{P}$ energies are
scaled by $N_{F}^{2}/2\Omega\lambda_{FB}\left(  \lambda_{FB}/\lambda
_{BB}\right)  $ and plotted as a function of $\alpha$ with mass ratio
$m_{B}/m_{F}=\frac{6}{7}$.}%
\label{ExEpEd}%
\end{figure}

We now list and describe the contributions of diagrams $(b)-(d)$. The diagram
$\left(  b\right)  $ describes fermion-mediated boson-boson interactions
\begin{equation}
\Delta E_{\left(  b\right)  }=\Delta E_{BB,\mathrm{fmed}}^{\left(  2\right)
}=-\frac{N_{B}^{2}}{2\Omega}\lambda_{FB}\alpha_{F}=-\frac{N_{B}^{2}}{2\Omega
}\lambda_{FB}\left(  1+\frac{m_{F}}{m_{B}}\right)  \left\vert \frac
{a_{BF}k_{F}}{\pi}\right\vert ,
\end{equation}
obtained earlier in Secs.\ \ref{Sec_linearresponse} and \ref{Sec_FmedBB}. In a
Hartree-Fock-like analysis of the boson-mediated fermion-fermion interactions,
diagrams $\left(  c\right)  $ and $\left(  d\right)  $ play a special role.
Diagram $\left(  c\right)  $ corresponds to the direct part of the
boson-mediated interaction,%
\begin{equation}
\Delta E_{\left(  c\right)  }=\Delta E_{FF,\mathrm{bmed},D}^{\left(  2\right)
}=-\frac{N_{F}^{2}}{2\Omega}\lambda_{FB}\left(  \frac{\lambda_{FB}}%
{\lambda_{BB}}\right)  ,
\end{equation}
and part of diagram $\left(  d\right)  $ gives the exchange interaction
contribution. Specifically, diagram $\left(  d\right)  $ corresponds to the
sum of the second-order polaron energy shift experienced by the fermions
caused by their interaction with the surrounding BEC phonon fluid and the
exchange part of the boson-mediated fermion-fermion interaction,
\begin{equation}
\Delta E_{\left(  d\right)  }=\Delta E_{FF,\mathrm{bmed},X}^{\left(  2\right)
}+\Delta E_{F,\mathrm{med},P}^{\left(  2\right)  }.
\end{equation}
In a static approximation of the boson-mediated interaction, the fermions
attract each other via an attractive Yukawa potential of range $\xi$. While
the static approximation is not always valid, the effective interaction does
have an effective range of order $\xi$ so that the $\alpha$ parameter
quantifies the ratio of the mediated interaction range to the average
fermion-fermion distance. The exchange interaction then depends on the
$\alpha$ parameter and the mass ratio (which affects the region in which the
static approximation is valid),
\begin{equation}
\Delta E_{FF,\mathrm{bmed},X}^{\left(  2\right)  }=\frac{N_{F}^{2}}{2\Omega
}\lambda_{FB}\left(  \frac{\lambda_{FB}}{\lambda_{BB}}\right)  e_{X}\left(
\alpha;\frac{m_{F}}{m_{B}}\right)  .
\end{equation}
The dimensionless exchange function $e_{X}(\alpha;m_{F}/m_{B})$ introduced in
the above equation satisfies $\lim\alpha\rightarrow0=1$, which ensures that
the exchange and direct contributions cancel in the limit that the interaction
range is much smaller than the average fermion-fermion distance, as required
by the Pauli principle. $\Delta E_{F,\mathrm{med},P}^{\left(  2\right)  }$
denotes the modification to the kinetic energy of ideal fermions with their
dispersion altered by the interaction with the surrounding BEC, as described
by second-order perturbation theory. This \textquotedblleft
polaron\textquotedblright\ contribution is often well described by a
zero-momentum energy shift and an effective mass value. For the contact
interaction, the second-order polaron contribution has to be renormalized. The
resulting energy shift experienced by a single fermion particle is of order
$\alpha_{B}\left(  \lambda_{FB}\rho_{B}^{0}\right)  $. The summation over
fermion occupation numbers gives a many-body contribution
\begin{equation}
\Delta E_{F,\mathrm{med},P}^{\left(  2\right)  }\simeq N_{F}\alpha_{B}\left(
\lambda_{FB}\rho_{B}^{0}\right)  =N_{B}\alpha_{B}\left(  \lambda_{FB}\rho
_{F}^{0}\right)  ,
\end{equation}
which can be rewritten in the energy units of the boson-mediated interactions
using $N_{B}\alpha_{B}=(3/2)\alpha^{-3}N_{F}\left(  \lambda_{FB}/\lambda
_{BB}\right)  $,
\begin{equation}
\Delta E_{F,\mathrm{med},P}^{\left(  2\right)  }=\frac{N_{F}^{2}}{2\Omega
}\lambda_{FB}\left(  \frac{\lambda_{FB}}{\lambda_{BB}}\right)  e_{P}\left(
\alpha;\frac{m_{F}}{m_{B}}\right)  ,
\end{equation}
where the remaining $\alpha$ dependence is absorbed by the dimensionless
polarization function $e_{P}$. Note that the scale of the boson-mediated
fermion-fermion interaction energy per fermion particle is $\lambda_{FB}%
\rho_{F}^{0}$,whereas the scale of the fermion-mediated boson-boson
interactions per boson particle is%
\begin{equation}
\lambda_{FB}\rho_{B}^{0}\alpha_{F}=\lambda_{FB}\rho_{B}^{0}\left(
1+\frac{m_{F}}{m_{B}}\right)  \left(  \frac{a_{FB}k_{F}}{\pi}\right)  .
\end{equation}
This also implies that the ratio of the boson-mediated fermion energy scale to
the fermion-mediated boson interaction energy is $\left[  \rho_{F}^{0}/\left(
\alpha_{F}\rho_{B}^{0}\right)  \right]  \left(  \lambda_{BF}/\lambda
_{BB}\right)  =(8/3\pi)\alpha^{2}$. In Fig.\ \ref{ExEpEd}, we show the
dimensionless polaron and exchange functions as a function of $\alpha$ for a
specific choice of the mass ration $m_{B}/m_{F}=6/7$ (corresponding to $^{7}%
$Li - $^{6}$Li mixtures). 

Below, we consider a specific region of the BEC-fermion mixture density space.
Starting with a homogeneous mixture of repelling particles $\left(
a_{BB}>0,a_{BF}>0\right)  $, comparable fermion and boson densities, and large
fermion-boson scattering length (i.e., $a_{BF}\gg a_{BB}$, while still
maintaining $a_{BF}k_{F}\ll1$), we gradually reduce the boson density. In
reducing the number of boson particles to zero, we cannot expect the
second-order description of this section to remain valid. A few boson
particles cannot mediate interactions for a large number of fermions and give
a boson-density independent mediated fermion-fermion interaction.
Nevertheless, in the process of decreasing the BEC density, $\alpha_{F}$
remains constant whereas $\xi$ increases as $\xi\thicksim(\rho_{B}^{0}%
)^{-1/2}$ so that $\alpha_{B}\rightarrow0$ and $\alpha\rightarrow\infty$ as
$\alpha=2\left(  m_{B}/m_{F}\right)  \left(  \alpha_{F}/\alpha_{B}\right)  $.
Notice that in this dilute BEC limit, the contributions of diagram $\left(
d\right)  $ vanish as $e_{P.}$ and $e_{X}$ rapidly tend to zero as $\alpha$
increases above unity.

We could have suspected the vanishing of the diagram $\left(  d\right)  $
contribution as its expression involves a summation over a boson momentum so
that we may expect at least a factor $\alpha_{B}$. However, the $\alpha$
dependence of the dimensionless integrals complicates the analysis.
Nevertheless, the numerical results shown in Fig.\ \ref{ExEpEd} confirm the
naive expectation of vanishing energy contribution in the dilute BEC limit.
This finding indicates a general trend, namely that those diagrams with
BEC-phonon momenta that are summed over vanish in the low-density BEC limit.
We can expect the diagrammatic analysis to simplify considerably in the dilute
BEC region. Even though we consider a low BEC density region, part of this
region displays interesting many-body behavior as the fermion-mediated
boson-boson attraction competes with the usual contact-interaction boson-boson repulsion.

\section{Summation of stretched, linear diagrams\label{Sec_sum}}

The general diagrammatic analysis becomes unwieldy as the values of the
coupling parameters increase. The resulting complexity is a common failing of
diagrammatic treatments, often limiting their usefulness in describing
strongly interacting systems. Fortunately, in the fermion-BEC mixture, a
significant simplification occurs in the low-density BEC limit. For that
system, we define the low BEC density regime as the limit in which the BEC is
dilute with respect to the usual (contact) boson-boson interactions,
$\sqrt{\rho_{B}^{0}a_{BB}^{3}}\ll1$, and in which the bosons do not
significantly affect the fermion particle properties (altering their effective
mass, for instance, or including vertex corrections in the description of
effective fermion-boson interactions). The latter conditions are satisfied if
\begin{equation}
\left(  1+\frac{m_{F}}{m_{B}}\right)  \left(  1+\frac{m_{B}}{m_{F}}\right)
\left(  \frac{a_{FB}}{a_{BB}}\right)  \frac{a_{FB}/\xi}{\pi}\ll
1\;.\label{lowd}%
\end{equation}
In the limit that the fermion-boson interaction is increased, $\left(
1+m_{F}/m_{B}\right)  (a_{FB}/a_{BB})\gg1$, condition Eq.\ (\ref{lowd}) is a
stronger requirement than $\alpha_{B}\ll1$. More generally, diagrams in which
phonon momenta have to be summed over can be discarded in this low-density BEC
limit. A large class of diagrams is still relevant, however, namely those
diagrams in which fermion \textquotedblleft bubbles\textquotedblright%
\ (particle-hole pairs) are connected by phonon propagators with momenta of
vanishing value in the long-wavelength limit implied in loops. This is the
case for the diagrams that we will refer to as \textquotedblleft
linear\textquotedblright\ diagrams--diagrams with two loops in which one can
move linearly from one loop to the other in only one way, crossing all the
interaction vertex signs ($\times$) while passing alternating fermion-bubble
and phonon segments.

\begin{figure}[ptb]
\begin{center}
\includegraphics[height=1.34in, width=3.5in] {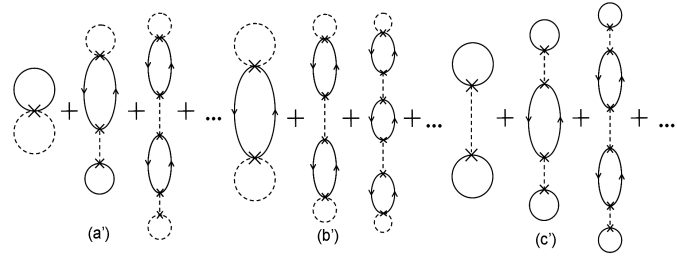}
\end{center}
\caption{Higher-order diagrams that are summed.}%
\label{highdiagram}%
\end{figure}

\begin{figure}[ptb]
\begin{center}
\includegraphics[width=3.4653in,height=2.5555in]{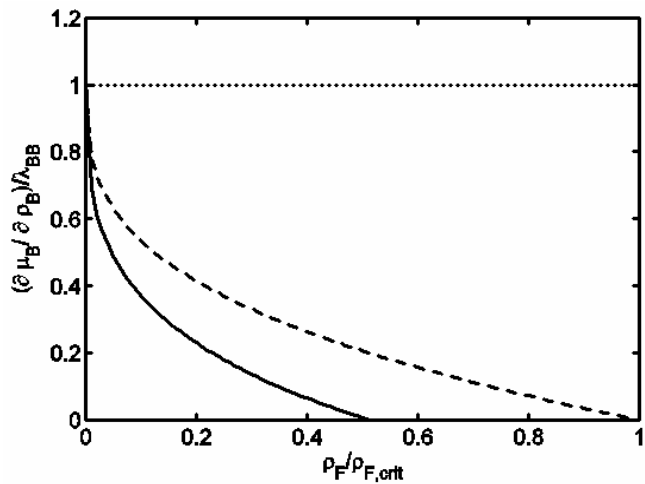}
\end{center}
\caption{$\left(  \partial\mu_{B}/\partial\rho_{B}\right)  $ (scaled with
$\lambda_{BB}T$) is plotted as a function of $\rho_{F}$ (scaled with
$\rho_{B,crit}$) for mean-field approximation (dotted line), second order
perturbation (dashed line), and infinite summation (solid line).}%
\label{dmu/drho}%
\end{figure}

An important subclass of the linear diagrams is the 'stretched' diagrams in
which the loops are placed at the earliest and at the latest time and in which
the bubble-phonon segments proceed from the earliest to the latest time in a
time-ordered fashion, as shown in Fig.\ \ref{highdiagram}. The $\left(  a^{\prime}\right)  $, $\left(  b^{\prime}\right)  $,
and $\left(  c^{\prime}\right)  $ diagram series shown in that graph follow
from Sec.\ \ref{Sec_cluster}, $\left(  a\right)  $, $\left(  b\right)  $, and
$\left(  c\right)  $ diagrams shown in Fig.\ \ref{diagram}, by including
larger and larger numbers of bubble-phonon insertions in between the loops.
Mathematically, each of the next insertions introduces an extra factor
$z=\left(  \alpha_{F}/4\right)  $ $\left(  \lambda_{BF}/\lambda_{BB}\right)  $
to the previous term. The $z$ parameter quantifies the magnitude of
fermion-mediated boson-boson interactions relative to the usual short-range
boson-boson interactions. In the infinite series summation over stretched
linear diagram $z$ takes on the role of an expansion parameter,%
\begin{align}
z &  =\frac{\alpha_{F}}{4}\frac{\lambda_{BF}}{\lambda_{BB}}\\
&  =\left(  1+\frac{m_{F}}{m_{B}}\right)  \left(  1+\frac{m_{B}}{m_{F}%
}\right)  \frac{a_{BF}k_{F}}{8\pi}\frac{a_{BF}}{a_{BB}}.
\end{align}
The series portrayed in $\left(  a^{\prime}\right)  $, $\left(  b^{\prime
}\right)  $, and $\left(  c^{\prime}\right)  $ take on the form $A\left(
1+z+z^{2}+\cdots\right)  =A/\left(  1-z\right)  $, yielding an equation of
state equal to
\begin{equation}
E=\frac{\lambda_{BB}N_{B}^{2}}{2\Omega}\left(  \frac{1-5z}{1-z}\right)
-\frac{\lambda_{BF}N_{F}^{2}}{2\Omega}\frac{\lambda_{BF}}{\lambda_{BB}}\left(
\frac{1}{1-z}\right)  +\frac{\lambda_{BF}N_{F}N_{B}}{2\Omega}\left(
\frac{1+z}{1-z}\right)  +\frac{3}{5}N_{F}\frac{\hbar^{2}k_{F}^{2}}{2m_{F}%
},\label{eos}%
\end{equation}
where in accordance to the low BEC density limit we have omitted the depletion
contribution $\left(  \lambda_{BF}N_{F}^{2}/2\Omega\right)  \sqrt{\rho
_{B}a_{BB}^{3}}$ to the zeroth-order energy $E_{0}$. The second-order analysis
resulted in a mechanical instability that corresponds to a diverging boson
compressibility at a fermion density equal to $z=\frac{1}{4}$ in terms of the
new expansion parameter. The present series of stretched linear diagram
summation carried out in the low-density BEC limit further shrinks the
mechanically stable region to $z<\frac{1}{5}$ as the equation of state of
Eq.~(\ref{eos}) implies that $\partial\mu_{B}/\partial\rho_{B}=\lambda
_{BB}\left(  1-5z\right)  /\left(  10z\right)  .$ In Fig.\ \ref{dmu/drho}, we
show the boson density derivative of the BEC chemical potential, proportional
to the inverse of the boson compressibility, as a function of the fermion
density. The density unit on the horizontal axis is $\rho_{F,\mathrm{crit}}$,
the critical fermion density calculated in the mean-field and second-order
analysis. In the mean-field equation of state, the derivative $\partial\mu
_{B}/\partial\rho_{B}$ does not depend on the fermion density as this
description neglects the fermion-mediated interactions. This result is
qualitatively wrong, and in the second-order analysis the derivative does
depend on the fermion density but not on the BEC density. Its value vanishes
at the fermion density $\rho_{F,\mathrm{crit}}$, corresponding to $z=\frac
{1}{4}$. Our low BEC density limit analysis in which we sum over the stretched
linear diagrams gives the lowest curve shown in Fig.\ \ref{dmu/drho}, still
depending only on the fermion density and vanishing at a fermion density
$\rho_{F,\mathrm{stret}}$ equal to
\begin{equation}
\rho_{F,\mathrm{stret}}=\left(  \frac{4}{5}\right)  ^{3}\rho_{F,\mathrm{crit}%
}\simeq0.5\rho_{F,\mathrm{crit}},
\end{equation}
roughly half of the previous value. We expect that a factor of $2$ difference
in density should be measurable in cold atom traps.

However, the stretched diagram analysis does not provide a definitive answer
even in the low-density BEC limit. So far, we have neglected linear diagrams.
The missing Goldstone diagrams can be obtained geometrically by folding the
stretched diagrams so as to have time intervals in which multiple excitations
occur simultaneously. To judge their importance we calculate the lowest-order
folded linear diagram contributions--the corresponding diagrams are shown in
Fig.\ \ref{diagram}. Their calculation is more involved as some propagators
now contain multiple excitation energies. Their combined contribution is equal
to
\begin{equation}
\Delta E_{3,\alpha}+\Delta E_{3,\beta}=\lambda_{BF}\frac{N_{F}N_{B}}{\Omega
}2z\left[  1+\left(  1-\frac{c}{v_{F}}\right)  \ln\left(  1+\frac{v_{F}}%
{c}\right)  \right]  \;.
\end{equation}
The simultaneity of phonon and fermion-particle-hole excitations in the
diagrams gives rise to a dependence on the velocity ration $r=c/v_{F}%
=m_{F}/\left(  2m_{B}\alpha\right)  $. Note that the dependence on fermion and
boson numbers is the same as that of the corresponding stretched diagram and
so is the order in $z$, as well as the sign of the contribution (the latter
only depends on the order of the diagram).

We expect that the main contributions to the BEC compressibility value stem
from the diagrams $\left(  b^{\prime}\right)  $ of Fig.\ \ref{highdiagram} and
their folded versions. The lowest-order folded $\left(  b^{\prime}\right)  $
diagrams are folded version of the second diagram of the $\left(  b^{\prime
}\right)  $ series--the diagram that has two boson loops, two fermion
particle-hole bubbles and one phonon propagator. These fourth-order folded
diagrams contribute negative terms to the equation of state and would tend to
further lower the value of the fermion density of diverging boson
compressibility below $\rho_{F,\mathrm{stret}}$. As their stretched
counterpart, they contribute terms that are proportional to $N_{B}^{2}$ and to
$z^{2}$. Unlike the stretched linear diagram series, these contributions will
also depend on the velocity ratio $r=c/v_{F}$. Hence, we expect them to
correct the boson compressibility at order $z^{2}$ and higher. The corrections
will then give a boson compressibility that depends on not only the fermion
density but also on the BEC density. Actually, at very low values of the boson
density, this dependence may be significantly affected by other terms as well.
Because of the folded diagram $r$ dependence, the third-order diagrams shown
in Fig.\ \ref{highdiagram} contribute a term to $\partial\mu_{B}/\partial
\rho_{B}$ that, for low values of $r$, $r\ll1$, varies as $\lambda_{FB}\left(
\rho_{B}/\rho_{F}\right)  zr\left[  1-\ln\left(  r\right)  \right]  $. At very
low boson densities this term would, in fact, dominate, but since $\left(
\rho_{F}/\rho_{B}\right)  \left(  c/v_{F}\right)  =\sqrt{\left(
8/3\pi\right)  \left(  a_{BB}k_{F}\right)  \left(  \rho_{F}/\rho_{B}\right)
}$, under most experimental conditions, this would happen at BEC densities
that would be so low it would be challenging to image their
profiles.

\section{Conclusions}

In this paper, we have developed a perturbation treatment of the
zero-temperature equation of state of single spin fermion-boson mixtures with
particles that interact via short-range interactions. This study reveals
qualitative failings of the mean-field equation of state. For instance, the
mechanical instability associated with the spinodal decomposition of the
phase-separation transition, which showed up as a saddle-point instability in
the mean-field description, turned out to be a point of diverging boson
compressibility. A first- and second-order calculations of the equation of
state show that the fermion-mediated boson-boson attraction overcomes the
short-range boson-boson repulsion at the same value for the fermion density,
$\rho_{F,\mathrm{crit}}$, at which the saddle-point instability occurred in
the mean-field analysis. We then showed that in the low BEC density limit, the
diagrammatic analysis simplifies to a study of the linear diagrams, of which
the \textquotedblleft stretched\textquotedblright\ diagram subclass can be
summed, giving an expansion parameter $z$ that quantifies the magnitude of the
long-wavelength fermion-mediated interaction relative to the regular
short-range boson-boson interactions. The resulting equation of state yields a
value of the fermion density at which the boson compressibility diverges that
is slightly higher than half corresponding to a factor $(4/5)^{3}$ of
$\rho_{F,\mathrm{crit}}$. A further exploration of the remaining linear
diagrams shows that the \textquotedblleft folded\textquotedblright\ diagrams
contribute terms that are higher order in $z$, and we speculate that these
contributions will further lower the fermion density at which the mechanical
phase-separation instability would set in, giving a critical value of the
fermion density that would now depend on the boson density as well. In
addition to revealing the inadequacy of the mean-field description near phase
separation, these studies have also identified the fermion-mediated
interaction as the cause of the mechanical instability associated with the
spinodal decomposition of the fermion-boson mixtures. In a general context,
the study presented in this paper gives a concrete illustration of how
correlations and how the competition of weak interactions can become important
near quantum phase transitions in the presence of interactions that, measured
by absolute standards, remain weak. The results also suggest that careful
experimental measurements near the phase separation transition of
fermion-boson mixtures can explore fundamentally interesting quantum many-body behavior.

\begin{acknowledgments}
This work was partly funded by the Laboratory Directed Research and
Development (LDRD) program of Los Alamos National Laboratory.
\end{acknowledgments}


\begin{thebibliography}{99}                                                                                               %


\bibitem {Kett02}Z. Hadzibabic, C. A. Stan, K. Dieckmann, S. Gupta, M. W.
Zwierlein, A. Gorlitz, and W. Ketterle, Phys. Rev. Lett. \textbf{88}, 160401 (2002).

\bibitem {Hulet03}K. E. Strecker, G. B. Partridge, and R. G. Hulet, Phys. Rev.
Lett. \textbf{91}, 080406 (2003).

\bibitem {Salomon01}F. Schreck, L. Khaykovich, K. L. Corwin, G. Ferrari, T.
Bourdel, J. Cubizolles, and C. Salomon, Phys. Rev. Lett. \textbf{87}, 080403 (2001).

\bibitem {Fernando}F. M. Cucchietti and E. Timmermans, Phys. Rev. Lett.
\textbf{96}, 210401 (2006).

\bibitem {Viverit}L. Viverit, C. J. Pethick, H. Smith, Phys. Rev. A
\textbf{61}, 053605 (2000).

\bibitem {Pinesfirst}D. Pines, \textit{Liquid Helium} (Academic, New York, 1963).

\bibitem {Cohen}J. M. J. van Leeuwen and E. G. D. Cohen, in
\textit{Proceedings of the Eighth International Conference on Low-Temperature
Physics} - edited by R. O. Davies (Butterworths Scientific Publications Ltd.,
London, 1963).

\bibitem {sep}For an overview of the low temperature helium mixture studies
and our current understanding of their many-body structure see: E. Krotscheck
and M. Saarela, Phys. Rep. \textbf{232}, 1 (1993).

\bibitem {Dima}D. Mozyrsky, I. Martin, and E. Timmermans, Phys. Rev. A
\textbf{76}, 051601(R) (2007).

\bibitem {Lifshitz}I. M. Lifshitz and Y. Kagan, Y, Soviet Physics, JETP,
\textbf{35}, 206, (1972).

\bibitem {Bardeen}J. Bardeen, G. Baym, and D. Pines, Phys. Rev. Lett.
\textbf{17}, 372 (1966).

\bibitem {Bardeen2}J. Bardeen, G. Baym, and D. Pines, Phys. Rev. \textbf{156},
207 (1967).

\bibitem {Wilkens03}A.P. Albus, F. Illuminati, and M. Wilkens, Phys. Rev. A
\textbf{67}, 063606 (2003).

\bibitem {Wilkens02}A.P. Albus, S.A. Gardiner, F. Illuminati, and M. Wilkens,
Phys. Rev. A \textbf{65}, 053607 (2002)

\bibitem {Viverit2}L. Viverit and S. Giorgini, Phys. Rev. A \textbf{66},
063604 (2002).

\bibitem {Eftemov}D. V. Efremov and L. Viverit, Phys. Rev. B., \textbf{65},
134519 (2002).

\bibitem {Bulgac}A. Bulgac, M. McNeilForbes, and A. Schwenk, Phys. Rev. Lett.,
\textbf{97}, 020402 (2006).

\bibitem {Timmermans}E. Timmermans, Phys. Rev. Lett. \textbf{87}, 240403 (2001).

\bibitem {Molmer}K. M\"{o}lmer, Phys. Rev. Lett. \textbf{80}, 1804 (1998).

\bibitem {Roth}R. Roth, Phys. Rev. A \textbf{66}, 013614 (2002)

\bibitem {collapse}G. Modugno, G. Roati, F. Riboli, F. Ferlaino, R. J. Brecha,
and M. Inguscio, Science, \textbf{297}, 2240 (2002).

\bibitem {HanPu}Han Pu, Weiping Zhang, M. Wilkens, and P. Meystre, Phys. Rev.
Lett. \textbf{88}, 070408 (2001).

\bibitem {Yip}S. K. Yip, Phys. Rev. A \textbf{64}, 023609 (2001).

\bibitem {Deborah1}D. H. Santamore, S. Gaudio, and E. Timmermans, Phys. Rev.
Lett. \textbf{93}, 250402 (2004).

\bibitem {Deborah2}D. H. Santamore and Eddy Timmermans, Phys. Rev. A
\textbf{72}, 053601 (2005).

\bibitem {Pines}A. Miller and D. Pines, Phys. Rev. \textbf{127}, (5) 1452 (1962).

\bibitem {Gold}J Goldstone, Proc. R. Soc., \textbf{A 239}, 267 (1957).
\end{thebibliography}
\end{document}